\documentclass[manuscript]{acmart}
\usepackage{xcolor}
\usepackage[draft,inline,nomargin,index]{fixme}
\usepackage{listings}
\usepackage{xspace} 
\usepackage{url}

\newcommand{\capture}{\textsc{Capture}\xspace}
\newcommand{\store}{\textsc{Store}\xspace}
\newcommand{\use}{\textsc{Use}\xspace}

\renewcommand\footnotetextcopyrightpermission[1]{} 
\begin{document}

\title{Audit Trails for Accountability in Large Language Models}

\author{Victor Ojewale}
\email{victor_ojewale@brown.edu}
\affiliation{%
  \institution{Department of Computer Science, Brown University}
  \city{Providence}
  \state{RI}
  \country{USA}
}

\author{Harini Suresh}
\affiliation{%
  \institution{Department of Computer Science, Brown University}
  \city{Providence}
  \state{RI}
  \country{USA}
}

\author{Suresh Venkatasubramanian}
\affiliation{%
  \institution{Department of Computer Science, Brown University}
  \city{Providence}
  \state{RI}
  \country{USA}
}

\renewcommand{\shortauthors}{Ojewale et al.}

\begin{abstract}
Large language models (LLMs) are increasingly embedded in consequential decisions across healthcare, finance, employment, and public services. Yet accountability remains fragile because process transparency is rarely recorded in a durable and reviewable form. We propose \emph{LLM audit trails} as a sociotechnical mechanism for continuous accountability. An audit trail is a chronological, tamper-evident, context-rich ledger of lifecycle events and decisions that links technical provenance (models, data, training and evaluation runs, deployments, monitoring) with governance records (approvals, waivers, and attestations), so organizations can reconstruct what changed, when, and who authorized it.  

This paper contributes: (1) a lifecycle framework that specifies event types, required metadata, and governance rationales; (2) a reference architecture with lightweight emitters, append only audit stores, and an auditor interface supporting cross organizational traceability; and (3) a reusable, open-source Python implementation that instantiates this audit layer in LLM workflows with minimal integration effort. We conclude by discussing limitations and directions for adoption.
\end{abstract}


\maketitle

\section{Introduction}
As Large language models have moved from research prototypes to infrastructure that shapes decisions in finance, healthcare, law, and public services\cite{nazi_large_2024, li_large_2023, lai_large_2024} it has become increasingly crucial that we have accountability and transparency mechanisms. When an error or harm occurs, organizations often cannot reconstruct which version was live, which data influenced it, who approved its release, or why particular changes were made. External audits and investigative reporting continue to surface harms \cite{buolamwini_gender_2018,Diakopoulos2016,Obermeyer2019}, yet the internal records needed to trace decisions are often fragmented across tools, stored in ad hoc ways, or never captured in the first place. The result is a persistent accountability gap.

Internal auditing frameworks argue for end to end oversight that is built into everyday practice \cite{Raji2020} and traceability has been proposed as a way to make accountability concrete in computing systems\cite{kroll2021}. Documentation artifacts such as model cards, datasheets, and FactSheets improve disclosure about models and datasets\cite{Mitchell2019,Gebru2018,Arnold2019}, but they remain snapshots at a point in time. MLOps practices improve technical provenance \cite{mlops2023,Vartak2016} yet often omit governance details such as who approved a change, under what conditions, and for what reasons. At the same time, AI development increasingly resembles a supply chain with many hands and limited visibility across organizational boundaries \cite{Thompson1980Responsibility,Widder2023Modularity,Cobbe2023SupplyChain,Bommasani2024Ecosystem,hopkins2025aisupplychainsemerging}, which makes accountability hard to assign without a shared, time-stamped record of changes and approvals across the development lifecycle. 

We address this gap by proposing \emph{LLM audit trails} as a way for organizations to embed accountability into the development and operation of LLM based systems. An audit trail is a chronological, tamper evident, and context rich ledger of events and decisions across selection, data use, fine tuning, evaluation, deployment, monitoring, and incident response. Rather than customizing logging for each pipeline, we conceptualize audit trails as a thin, system agnostic layer between heterogeneous LLM workflows and governance regimes. Audit trails turn transparency by design from principle into operational practice \cite{Felzmann2020} and align with regulatory record keeping \cite[Article 12]{euaiact} and risk management guidance \cite{NIST2023}.

Our contributions are:

\begin{enumerate}
  \item a lifecycle framework that specifies what to log and why across these stages, including the governance rationale for each stage;
  \item a reference architecture for capturing, storing, and using logs that emphasizes tamper evidence, privacy, and cross organizational linkage;
  \item a lightweight Python library that instruments common LLM development workflows to show that such trails are feasible with modest overhead.
\end{enumerate}

We motivate the need for audit trails through concrete scenarios, situate our proposal within work on accountability and documentation, present the framework and system design, demonstrate a proof of concept implementation, and revisit the scenarios to illustrate how audit trails can support investigation and governance in practice.

\section{Why do we need audit trails?}
\label{sec:motivation}

This section illustrates why audit trails are needed in practice by walking through two LLM deployment scenarios in finance, and healthcare. For each case we describe how the system is used in context, what the development and deployment choices look like in practice, what goes wrong, and what kinds of records investigators would need in order to understand and govern the system.

\paragraph{Financial advice chatbot incident.}

Retail banks are actively piloting generative AI for customer-facing assistance and internal analyst workflows, with early deployments focusing on chat-based support, document summarization, and other productivity tasks \cite{crisanto_regulating_2024,hsbc_mistral}

\emph{Scenario.}
A retail bank’s digital channels team deploys BankGPT, an LLM based assistant to answer mortgage questions for retail customers, while human loan officers retain formal authority over lending decisions. The assistant is positioned as a first line tool that can explain products, estimate affordability, and route customers to appropriate staff.

\emph{Operationalization.}
The bank’s ML team selects a proprietary base LLM from a provider and fine tunes it on historical customer service transcripts for mortgage support and bank mortgage policies, after a model risk committee approves the use of that base model and the fine tuning dataset. Engineers configure prompt templates, set the generation settings that control how long and how varied the assistant’s answers are, connect the model to internal product catalogs so it can look up current rates and eligibility rules, and define guardrail policies to block clearly inappropriate content. They then deploy the fine tuned model behind a feature flag to a subset of web and mobile users, with updates rolled out through the bank’s continuous integration and continuous deployment (CI/CD) pipeline.

\emph{Key event.}
Months later, a branch loan officer escalates a complaint from a customer who followed the assistant’s guidance and was later told by underwriting that the recommended mortgage product was unsuitable given their income and risk profile. The bank’s risk and compliance teams open an internal review to understand whether the LLM behaved within its intended scope and whether deployment controls were followed.

\emph{Governance needs.}
To investigate, the bank needs records that show which fine tuned model version was serving in the relevant channel at the time of the interaction, which prompt templates and decoding settings were active, and how rollout controls such as feature flags were configured. They also need to see what approvals governed that deployment, whether any emergency hotfixes were applied, and how those changes were authorized. In many current LLM deployments, such information is scattered across experiment trackers, CI logs, configuration files, and email threads, making it difficult to reconstruct a coherent timeline for internal review or external scrutiny.

\paragraph{Clinical documentation and follow up gaps.}

Clinical documentation remains a major source of administrative burden, with time-motion and log-based studies showing substantial clinician time devoted to EHR work and after-hours documentation \cite{sinsky_allocation_2016, arndt_tethered_2017}. In response, vendors and health systems are increasingly experimenting with “ambient” or note-drafting assistants that generate visit summaries and follow-up plans\cite{10.1001/jamanetworkopen.2025.34982}.

\emph{Scenario.}
An integrated health system deploys NoteAssist, an LLM based assistant that drafts visit summaries and suggested follow up plans for primary care clinicians. The assistant is framed as a documentation aid only, with clinicians retaining responsibility for verifying text and placing orders.

\emph{Operationalization.}
A vendor fine tunes a base LLM on de identified visit notes and discharge summaries, using supervised learning and clinician written examples to teach the model how to structure assessments and plans. The health system’s clinical leadership and informatics committee review evaluation reports on note quality and safety, approve a limited rollout in adult primary care, and document expectations that clinicians will review and edit all suggestions. The assistant is then integrated into the electronic health record so that when a visit is closed, NoteAssist generates a draft summary and follow up section that clinicians can accept, edit, or overwrite.

\emph{Key event.}
Several months after rollout, the patient safety office notices a cluster of incident reports and chart reviews in which recommended follow up for abnormal lab results or imaging findings was delayed. In many of these cases, the clinician used NoteAssist to draft the visit note. The safety and quality team opens a focused review to understand whether the assistant contributed to follow up gaps, for example by omitting certain recommendations, over generalising existing templates, or encouraging over reliance on default plans.

\emph{Governance needs.}
To conduct this review, the health system needs evidence about which version of NoteAssist was active during the affected encounters, which prompt templates and configuration settings were used for generating follow up plans, and how those settings changed over time. They also need to know how often clinicians accepted the assistant’s suggestions with minimal edits, how training and evaluation data reflected different clinical domains, and what approvals or risk assessments accompanied expansion from a small cohort of early adopters to wider use. In many current deployments, these details are spread across vendor documentation, EHR configuration, experiment tracking tools, and email threads, which makes it difficult to reconstruct a reliable, time aligned account of what the LLM was doing when the safety concerns arose. 

\section{Background and related work}\label{sec:background}

\paragraph{Accountability, traceability, and audits}
Accountability in sociotechnical systems requires identifiable actors, mechanisms through which they can be questioned and held responsible (for example internal review processes, regulators, or courts) and the possibility of consequences, all supported by records that make actions and justifications traceable \cite{10.1145/3351095.3372833,Diakopoulos2016}. External audits and investigative work have been central to revealing discriminatory outcomes and disparate performance in deployed AI systems \cite{buolamwini_gender_2018,mattu_how_nodate,Obermeyer2019}, but these efforts typically operate with limited visibility into internal processes and logs, motivating calls calls for internal, continuous auditing regimes
\cite{Raji2020,10516659,10.1145/3706598.3713301}. In parallel, legal and policy work has argued that traceability --  the ability to reconstruct which artifacts, configurations, and decisions produced a system’s behavior at a given time by linking outcomes back to design choices, data use, and responsible parties -- is a key mechanism for operationalizing accountability\cite{kroll2021,10.1145/3442188.3445921, 10.1145/3531146.3534628}.

For LLMs, Mokander et al argue for layered audit regimes that distinguish governance level, model level, and application level audits, with governance records (for example who approved a deployment, what risks were assessed or waived, what scope conditions were imposed) treated as first class evidence rather than informal background context \cite{Mokander2023}. This idea aligns with work hat highlight the limits of interpretability and behavioral probing for very large models \cite{bender2021,10.5555/3666122.3669397,Bommasani2021FoundationModels,10.5555/3666122.3668547} and emphasizes the importance of process transparency.

At the same time, standards and regulation are converging on traceability and record keeping as central requirements. The NIST AI Risk Management Framework identifies traceability as a core characteristic of trustworthy AI and calls for organizations to maintain records that support oversight and investigation \cite{NIST2023}. The EU AI Act, particularly Article~12, requires providers of high risk AI systems to implement automatic logging that allows regulators to verify whether systems operated in accordance with their intended purpose and applicable obligations \cite[Article 12]{euaiact}. Sectoral guidance and management system standards, along with emerging vendor offerings in model governance and monitoring, likewise emphasize lineage tracking, approvals, and change management \cite{AWSGovernance,DataRobot2022,fiddler_observability_2025}. 

Across these strands, there remains a gap between high level calls for traceability and the concrete, tamper evident, governance aware logging that organizations would need to actually demonstrate how LLMs are selected, fine tuned, deployed, and monitored in practice. Our work addresses this gap by specifying an event level framework and a concrete audit layer that organizations can adopt and adapt inside their own pipelines.

\paragraph*{Documentation and lifecycle provenance}
Documentation artifacts such as model cards, datasheets, AI FactSheets, and now system cards seek to standardize disclosures about capabilities, data, evaluation protocols, and limitations \cite{Mitchell2019,Gebru2018,Arnold2019,alsallakh2022system}. 
However, they are curated and static: they typically summarize the state of a model or system at a particular release point, often compiled manually from internal notes and dashboards.

In parallel, MLOps tooling and experiment tracking systems provide fine grained technical lineage by logging hyperparameters, datasets, code versions, and metrics for training runs and model artifacts \cite{mlops2023,Vartak2016,10.1145/3399579.3399867,weights_bias}. These systems are usually complemented by general purpose version control like Git-based workflows for code and configuration.
While they improve reproducibility, they are usually oriented toward developer needs, not governance.
Documentation efforts in evaluation, such as HELM style benchmark reporting, take a complementary approach. HELM (Holistic Evaluation of Language Models) reports model performance across many tasks and metrics and tracks how well different application scenarios and risks are covered by the chosen benchmarks \cite{Liang2023HELM}. The focus is on aggregate outcome summaries, such as accuracy, robustness, or fairness scores across a group of tasks, datasets, and scenarios, rather than on the internal decision processes that led an organization to select a particular model, configure it in a certain way, or approve it for deployment. 

Our work builds on MLOps and experiment-tracking telemetry but augments it with explicit decision logging, append only storage, and auditor oriented views. This makes governance-relevant provenance queryable and verifiable over time, and links technical changes to the approvals, waivers, and scope conditions under which they occurred. 

\paragraph*{Supply chains and distributed responsibility}
Contemporary AI development increasingly resembles an algorithmic supply chain in which foundation model providers, fine tuning vendors, platform operators, and downstream deployers share responsibility for a composite system \cite{Cobbe2023SupplyChain,hopkins2025aisupplychainsemerging}. This modularity can obscure responsibility rather than clarify it, since components do not always compose cleanly and failures emerge at the seams between organizations and subsystems \cite{Widder2023Modularity,hopkins2025aisupplychainsemerging}. Macro level mapping efforts such as Ecosystem Graphs seek to chart relationships among datasets, models, and applications across this landscape \cite{Bommasani2024Ecosystem}, but they generally do not expose the fine grained, time stamped records needed to reconstruct who changed what configuration, on which model version, with which approvals, at which client.

Our focus on audit trails is complementary. By specifying shared identifiers and event schemas that can be implemented by different actors in the chain, we aim to make it possible for organizations to construct traceable histories of selection, fine tuning, configuration, deployment, and monitoring across organizational boundaries, while respecting contractual and privacy constraints.

\section{Conceptual Framework for LLM Audit Trails}\label{sec:framework}

This section defines the conceptual building blocks of an LLM audit trail: what an audit trail is, what metadata and rationale must accompany it, and how events relate across lifecycle stages and organizations. We use this framework to specify what should be captured at each lifecycle stage and why. Section~\ref{sec:architecture} then maps these requirements onto a concrete \emph{capture, store, and use} architecture, with lightweight event emitters, append-only storage, and auditor-facing tools for querying and verification, and Section~\ref{sec:implementation} provides a python implementation that can be integrated in LLM workflows.

An \textbf{audit trail} for a large language model is a chronological, tamper-evident record of lifecycle events and governance decisions that links technical decisions (artifacts, data versions, configurations, and evaluations) with organizational accountability (owners, approvals, waivers, and scope constraints). 
An audit trail should be able to support traceability across organizational boundaries even with partial information, because no single organization controls the full LLM development lifecycle. 

In what follows, we describe the elements of an audit trail at each stage of the LLM development cycle, broken up into pretraining, adaptation, deployment, and operational monitoring. Across all these stages, the audit trail we propose has several key characteristics:

\begin{itemize}
\item \textbf{Comprehensive coverage.} Trails span all relevant stages of the lifecycle, from corpus construction and pretraining through downstream adaptation, deployment, and monitoring, and explicitly document handoffs between organizations.
\item \textbf{Chronological integrity.} Entries are time stamped and sequenced, providing a clear narrative of evolution that is indispensable for investigations and for understanding how decisions unfolded.
\item \textbf{Immutability.} Once recorded, entries are tamper evident or read only so that they can serve as reliable evidence. This applies both within single organizations and across federated trails that link providers and deployers.
\item \textbf{Accessibility for audit.} Information is stored in structured, queryable formats that support internal review and external oversight, with well defined permissions and interfaces for legitimate access.
\item \textbf{Decisions as first class objects.} Governance actions such as approvals, waivers, and attestations, along with their rationales and scope, are recorded alongside technical telemetry, not left in informal documents or email threads.
\end{itemize}

\subsection{Pretraining data and foundation model development}

For organizations that train or release foundation models, an LLM audit trail begins upstream with pretraining data curation and model development. Recent open models such as OLMo \cite{groeneveld2024olmoacceleratingsciencelanguage} document corpus composition, filtering pipelines, and training configurations in considerable detail; our framework can be seen as a way to standardize and operationalize such records as structured, machine readable audit events.

At a minimum, the trail should record the provenance and governance of each corpus component (sources, collection method, licensing and restrictions), the high-level data processing pipeline (deduplication and filtering policies), and the resulting dataset versions that enter training. It should also capture the training lineage for released checkpoints, including model family and version identifiers, training configuration sufficient to reproduce the run, and the evaluation record used to justify release, with references to the scripts and benchmark versions used.

Finally, pretraining governance should be logged alongside this technical provenance, including approvals to initiate large-scale training, waivers or exceptions for particular data sources, and release decisions that define which model variants are published under which licenses and intended-use statements. For an open model such as OLMo \citep{groeneveld2024olmoacceleratingsciencelanguage}, these records turn a narrative model report into a verifiable chain of events linking released weights to specific data mixtures, training runs, and governance checkpoints, which downstream actors can reference when composing their own audit trails.

\subsection{Base Model Selection and Preparation}

Downstream work begins by selecting a base model and establishing the conditions under which it will be adapted or deployed. Even when pretraining is external, selection is a governance commitment because it fixes inherited limitations, contractual constraints, and upstream dependencies that can change over time. The audit trail should therefore record a clear, reviewable starting point for the lifecycle that follows.

The audit trail should therefore bind the project to an unambiguous base model identity and provenance, including the model name and version or release identifier, the acquisition channel (API endpoint or artifact registry), and any license or policy restrictions that constrain use. Where available, the trail should reference the provider documentation used to make the decision (for example a model card or system card) and record the best-available scope intent at the time of selection, such as the target domain, anticipated use classes, excluded uses, and any risk flags that trigger additional evaluation.

This stage should also establish accountable ownership by recording the initiating team or role, the production owner, and any required authorization to proceed. If the downstream deployment context is not yet fixed, the trail should treat later contextualization as an explicit handoff event, recorded when a concrete application and user population are determined.

\subsection{Adaptation, evaluation, and release readiness}

After selection, organizations adapt the model to a task and scope, either by weight updates (fine tuning) or by configuration over a fixed model (prompting, retrieval, tool use, and guardrails). The audit trail should capture whichever pathway is used by linking versioned inputs to a versioned model artifact or configuration bundle, together with the evidence used to judge readiness.

For fine tuning, the trail should connect the training run to a specific dataset version and its governance basis, including provenance and any approvals that permitted its use \citep{Gebru2018}. If human feedback is part of the adaptation workflow (for example demonstrations, preference rankings, or red-teaming annotations), the trail should record the provenance of that feedback process and the intermediate artifacts it produces, such as reward models, preference datasets, and the subsequent training runs that consume them \citep{Ouyang2022}. For configuration-driven adaptation, the same traceability requirement applies: prompts, retrieval indexes, tool schemas, and safety policies should be treated as versioned assets whose changes are attributable to actors and time.

Evaluation bridges adaptation to deployment. The trail should record evaluation results and their scope so that later reviewers can determine what was tested, with which data or scenario suites, and what limitations were known at sign off. This aligns with the view that governance evidence should be treated as priority audit material alongside technical measures \citep{Mokander2023}. Finally, the trail should record the release-readiness decision itself, including who authorized deployment, for what intended scope, and with what conditions or waivers.

\subsection{Deployment and System Integration}
Once a model is approved, it moves into production deployment. Deployment can take many forms: embedding the model in an application, exposing it via an API, etc. and the responsibility may lie with a systems integrator, customer organization, or platform provider. The audit trail at this stage should make the deployed system state reconstructible, so that reviewers can determine what was live, under what constraints, and under whose authority.

The trail should bind the approved model version (or approved configuration bundle) to a specific deployment instance, including when and where it was deployed and which surrounding service or application version invoked it. It should also capture the runtime configuration that materially shapes behavior in production, including prompt policies, decoding settings, retrieval or tool-use configuration, and any safety or guardrail layers. The goal is not to preserve every low-level knob, but to retain enough information to understand the context of behavior and to identify when changes in configuration plausibly account for changes in outcomes.

Finally, because deployments are governed through control and scope, the trail should record who can access and modify the deployed system (including rollback authority), as well as the go-live decision and any rollout strategy or scope constraints such as pilots, staged releases, or conditional approvals tied to user groups or geographies. These records let an auditor reconstruct whether a contested outcome occurred during a limited pilot, a partial rollout, or full deployment, and whether the system was operating within the approved scope at the time.

\subsection{Operational monitoring, feedback, and incident response}
After deployment, the audit trail continues to accumulate entries during the operation of the model. This post-deployment phase often spans long time horizons and multiple actors: deployers operate the system in context, users experience its outputs, and upstream providers or vendors may continue to patch, retrain, or change dependencies. The audit trail should therefore support two linked goals: ongoing oversight (is the system behaving within approved bounds) and retrospective investigation (what was the system, and what changed, when a harm or failure is alleged).

Operational logging should make usage and exposure legible while remaining compatible with privacy and scale constraints. The trail should record the logging posture and retention policy, and capture sufficient interaction metadata to establish when the system was used, at what scale, and under which deployment scope, with references to richer internal logs where applicable. Where model outputs influence downstream actions, the trail should also link outputs to decision points in a structured way so that reviewers can assess reliance and error propagation.

Finally, ongoing monitoring should appear as periodic evidence: records of recurring checks, evaluation refreshes, and signals of drift or abnormal guardrail activity while the trail should treat feedback, incidents, and remediation as important operational events. When concerns are raised by users, internal monitors, or external stakeholders, the trail should record when the issue was detected, who triaged it, what evidence was consulted, what decision was taken (for example rollback, patch, new guardrail, temporary suspension), and what follow-up obligations were created. The same logic applies to model updates: redeployments should be traceable to a documented reason, to an updated evaluation record, and to an explicit approval that links the new deployed version back to the previous operational history. Together, these operational records support both continuous accountability and credible post hoc explanations of how the deployed system evolved over time.

\begin{figure*}[t]
  \centering
  \includegraphics[width=\textwidth]{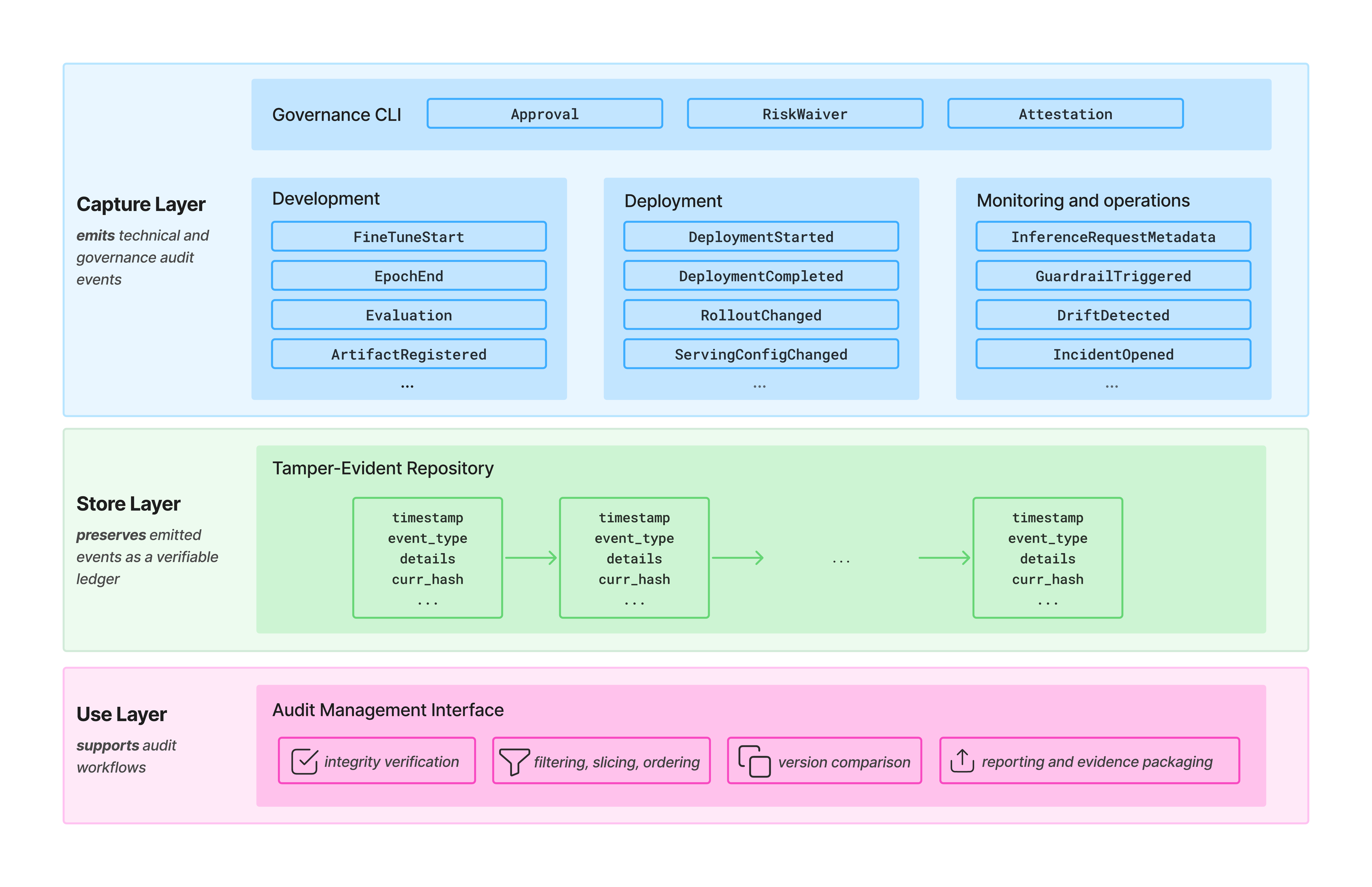}
   \Description{Proposed System Architecture for Audit trails as a reusable layer. Capture emits audit events(both technical and governance); Store preserves them as a verifiable ledger; Use supports audit workflows.}
  \caption{Proposed System Architecture for Audit trails as a reusable layer.}
  \label{fig:figma}
\end{figure*}

\section{Proposed system architecture}\label{sec:architecture}

We now describe a system architecture (see Fig. \ref{fig:figma}) to achieve the goals of an audit trail as set out above. The architecture is comprised of three layers: \capture (produce auditable events from technical workflows and governance checkpoints), \store (preserve those events in an append-only, tamper-evident trail), and \use (support auditor-facing querying, verification, and reporting). The architecture operates as a shared layer in the LLM stack rather than as a bespoke solution for any individual pipeline. 

\subsection{\capture layer: An emitter-based architecture}

\capture is built as a collection of \emph{emitters}. Intuitively, each event in the lifecycle of an LLM build and deployment emits data that can be logged. Emitters output data in a structured format with stable identifiers (for example \texttt{model\_id}, \texttt{dataset\_id}, \texttt{deployment\_id}) so that data from different parts of the lifecycle can be integrated in the next phases of the audit. 
Capture has two complementary sources.

\paragraph{Technical event sources.}
Here, we instrument existing tooling to emit machine-authored events of different kinds
\begin{itemize}
  \item \textbf{Development (training and evaluation).} Training framework callbacks, experiment trackers, and job wrappers record events such as \texttt{FineTuneStart}, \texttt{EpochEnd}, \texttt{Evaluation}, 
  
  \texttt{CheckpointSaved}, and \texttt{ArtifactRegistered}. These events attach technical provenance such as code version (for example Git commit), dataset version, hyperparameters, metrics, and model checksums.
  \item \textbf{Deployment.} CI/CD steps, model registries, and serving configuration managers record \texttt{DeploymentStarted}, 
  
  \texttt{DeploymentCompleted}, \texttt{RolloutChanged}, and 
  
  \texttt{ServingConfigChanged}. These events capture the runtime configuration that materially affects behavior, such as prompt template identifiers, decoding parameters, retrieval or tool-calling configuration, feature flags, environment, and rollback actions.
  \item \textbf{Monitoring and operations.} Service middleware and guardrail systems record \texttt{InferenceRequestMetadata}, 
  
  \texttt{InferenceResponseMetadata}, \texttt{GuardrailTriggered}, 
  
  \texttt{DriftDetected}, and \texttt{IncidentOpened/Resolved}. To manage privacy and scale, these events can log hashes and redacted summaries while pointing to richer internal logs governed by stricter access controls.
\end{itemize}

\paragraph{Governance checkpoints.}
Many lifecycle transitions depend on human judgment rather than automated tooling. The capture layer therefore includes a governance checkpoint mechanism that records structured, human-authored decision events:

\begin{itemize}
  \item \textbf{Approval} events for decisions such as authorizing a dataset for use, approving a model version for deployment, or expanding rollout scope.
  \item \textbf{RiskWaiver} events for deviations from standard controls under explicit time-bound or scope-bound conditions.
  \item \textbf{Attestation} events for statements about licensing, data scope, evaluation completeness, or policy compliance.
\end{itemize}

Each decision event records the decision owner, rationale or statement, the scoped identifiers it applies to, any constraints (for example permitted jurisdictions or user groups), and references to supporting artifacts such as tickets, risk assessments, or evaluation reports. Treating governance checkpoints as first-class events makes it possible to audit not only what changed technically, but also who authorized the change and under what conditions.

\subsection{\store layer: append-only, tamper-evident trails}
Emitted events need to be stored in a shared structure as a sequential trail that can be examined later. \store performs this function by storing events emitted in the Capture phase in an append-only trail. Each record can include an RFC3339 timestamp, event type, scoped identifiers, a \texttt{details} payload, and integrity metadata that makes in-place modification or deletion detectable. A simple mechanism is a hash chain that links each event to the previous event (\texttt{prev\_hash} $\rightarrow$ \texttt{curr\_hash}). Stronger deployments can add cryptographic signatures per writer or per organization to support multi-writer provenance.

The same design can be implemented over different storage backends depending on an organization’s constraints. In some settings, the trail may live in a managed database configured with immutability controls and write-once retention policies. In others, it may be stored as append-only files whose integrity is strengthened by anchoring to external timestamping or transparency services. In multi-party supply chains, a further option is a dedicated ledger operated by a neutral party that aggregates events from multiple organizations while preserving provenance across writers.

\paragraph{Inter-organizational usefulness without full sharing.}
In multi-party supply chains, organizations may be unwilling or unable to share raw logs, especially for pretraining data or sensitive operational traces. \store can still support cross-organizational traceability through \emph{signed pointers} and \emph{summaries}. Concretely, a party can publish a signed reference (\texttt{store\_uri}, \texttt{log\_id}, \texttt{event\_id}) and a small set of non-sensitive fields (for example model version identifiers, hashes of artifacts, and high-level attestations), enabling downstream auditors to verify linkage and integrity without access to underlying proprietary content. This supports composition of evidence across organizations while respecting contractual, privacy, and trade secret constraints.

\subsection{\use layer: auditor-facing querying, verification, and reporting}
\use is not a separate logging mechanism. It is the set of auditor-facing functions that operate over the stored trail produced by the capture and store layers. Because events are recorded in a common schema with stable identifiers and preserved in an integrity-protected, append-only log, the same underlying record can support multiple oversight tasks without requiring bespoke instrumentation for each one.

First, \use enables \textbf{integrity verification}. Auditors can replay the hash chain (and, where present, check signatures) to confirm that the trail has not been altered or truncated since it was written. This establishes the basic evidentiary property of the record: that the timeline being reviewed is complete relative to the stored log and that tampering would be detectable.

Second, \use supports \textbf{scoped reconstruction and comparison}. Because events are keyed by identifiers such as \texttt{model\_id}, \texttt{dataset\_id}, and \texttt{deployment\_id}, an auditor can filter and order the append-only stream to reconstruct a time-aligned timeline for a given model or deployment, and then group those slices to compare versions. This makes it possible to answer questions like which evaluated artifact was approved for production, what configuration was active during a given period, whether a deployment change occurred before or after an approval or waiver, and what materially changed between two approved releases (for example prompt templates, decoding parameters, retrieval settings, guardrail policies, or rollout scope), using the recorded timestamps and event details rather than scattered tickets or recollection.

Finally, \use enables \textbf{reporting and evidence packaging}. Since the trail is structured and integrity-protected, organizations can export a bounded subset of events with the accompanying integrity metadata (hash links and any signatures) for internal review, incident investigations, or external submissions. Access to these exports can itself be logged as additional events, creating a secondary record of who inspected which evidence and when.

In summary, \use is the ``read path'' for audit trails: it consumes the stored, chained event ledger and exposes verification, filtering, comparison, and reporting workflows that make governance questions answerable from the same underlying record.

\subsection{Interoperability and schema}

All events should share a compact core schema:
\texttt{\{event\_id, timestamp, system, actor, event\_type, model\_id, dataset\_id, 
deployment\_id, details, prev\_hash, curr\_hash, sig?\}}.
This core can be expressed as JSON Schema and extended through \emph{sector-specific profiles} that specify additional required fields and constraints. For example, a high-risk profile might require an \texttt{Approval} event in scope before any \texttt{DeploymentCompleted} event, or require periodic drift checks to be recorded at a specified cadence. 

This design feature is important. A shared core plus profiles supports diverse governance regimes without redefining logging for each system.

\section{Proof of Concept Implementation: \texttt{llm-audit-trail} Python Library}\label{sec:implementation}

We implemented a small open source Python library, \texttt{llm-audit-trail}, to demonstrate that the reference architecture can be integrated into common downstream LLM workflows with minimal code changes.\footnote{Source code: \url{https://github.com/victorojewale/audit-trail-PoC}} The Proof of Concept (PoC) targets three goals: (i) low-friction capture of lifecycle events, (ii) tamper-evident integrity via hash chaining, and (iii) explicit recording of human governance decisions alongside technical telemetry. The PoC is intentionally not a full governance platform and it implements a representative subset of the reference architecture from Section~\ref{sec:architecture} to illustrate feasibility. It implements \capture through lightweight integrations that emit structured events, \store through a hash-chained append-only JSONL ledger, and \use through a verifier and the possibility for simple filtering utilities that allow an auditor to reconstruct timelines for a model or deployment scope.

\subsection{\capture in the PoC: integrations and event emission}

We use existing callback methods within standard AI-stack components to implement  \capture emitters. This illustrates how the emitter-based framework can be easily integrated into existing frameworks.

\begin{figure}[t]
    \centering
    \begin{minipage}[t]{0.43\linewidth}
        \centering
        \includegraphics[width=\linewidth]{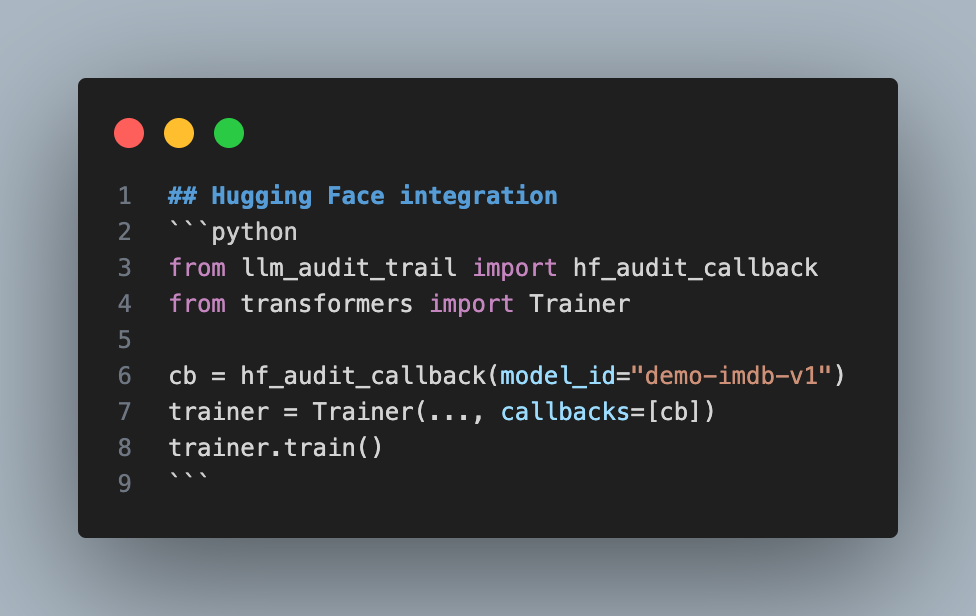}
        \Description{Screenshot showing the audit trail callback in a Hugging Face training script.}
        \caption{Minimal integration: attaching the audit trail callback in a Hugging Face training script.}
        \label{fig:audit-callback-snapshot}
    \end{minipage}
    \hfill
    \begin{minipage}[t]{0.48\linewidth}
        \centering
        \includegraphics[width=\linewidth]{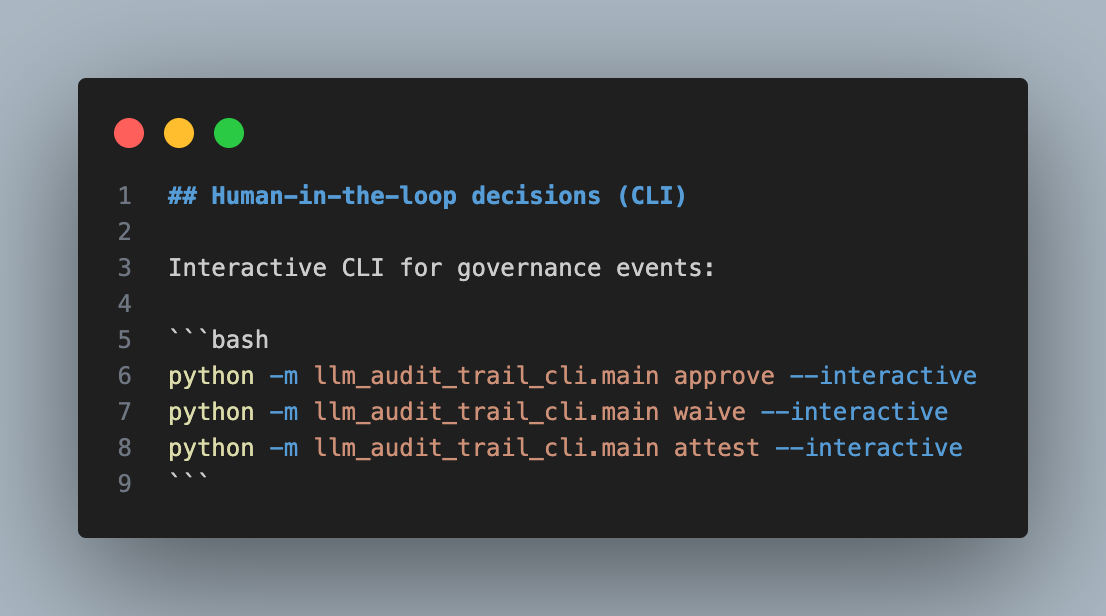}
        \Description{Screenshot of a terminal showing the governance CLI command to log an approval event in the audit trail.}
        \caption{Recording an approval decision via the governance CLI.}  \label{fig:cli-approve}
    \end{minipage}
\end{figure}

\paragraph{Training and evaluation emitter.}
A Hugging Face \texttt{TrainerCallback} (\texttt{AuditTrailCallback})(see Fig. \ref{fig:audit-callback-snapshot}) emits events such as \texttt{FineTuneStart}, \texttt{EpochEnd}, \texttt{Evaluation}, \texttt{CheckpointSaved}, and \texttt{FineTuneEnd}. Each event records run configuration (for example learning rate, batch size, epochs, seed, output directory) and the latest metrics (for example loss or accuracy), keyed by a stable \texttt{model\_id}. Users attach the callback by passing \texttt{callbacks=[hf\_audit\_callback(model\_id=...)]} to \texttt{Trainer}.

\paragraph{Dataset registration emitter.}
A helper function \texttt{register\_dataset} in the code base emits \texttt{DatasetRegistered} events that record dataset identifier, version, origin (for example a Hugging Face dataset URL), row counts, preprocessing summary, and optional content hashes or datasheet references. These events anchor later training and evaluation events to specific data versions.

\paragraph{Serving and monitoring emitter.}
A FastAPI middleware (\texttt{AuditMiddleware}) emits \texttt{InferenceRequestMetadata} and \texttt{InferenceResponseMetadata} events. It records request path and method, request and response hashes, latency, and an optional short preview that can be redacted by default. This demonstrates how deployment-time behavior can be mapped into the same audit surface while limiting sensitive content.

\paragraph{Script-level emitter.}
A generic \texttt{AuditLogger} can be used directly in scripts or CI jobs to emit ad hoc events such as \texttt{ModelDeployed}, \texttt{ServingConfigChanged}, or \texttt{IncidentOpened}. This supports workflows that do not fit a single framework integration.

\paragraph{Governance capture via CLI.}

The PoC includes a command-line tool \texttt{llm\_audit\_trail\_cli} that writes human-authored events into the same log. It supports three event types mentioned in the proposed architecture.

Each command (\texttt{approve}, \texttt{waive}, \texttt{attest}) can run interactively or via flags, prompting for owner, rationale or statement, scope identifiers, optional constraints, and references(see Fig. \ref{fig:cli-approve}). Prompts are driven by a small YAML configuration file that specifies required fields and labels, and the CLI can suggest recent identifiers by scanning the existing log.

\subsection{\store in the PoC: append-only JSONL and hash chaining}

The PoC store is a newline-delimited JSON log (\texttt{audit\_trail.jsonl}). Each entry is a single event record with a unique \texttt{event\_id}, an RFC3339 \texttt{timestamp}, scoped identifiers (\texttt{model\_id}, \texttt{dataset\_id}, \texttt{deployment\_id}), a free-form \texttt{details} payload, and integrity metadata (\texttt{prev\_hash}, \texttt{curr\_hash}).

The \texttt{curr\_hash} is a SHA-256 digest over the serialized event payload plus the previous hash. The first event uses a fixed \texttt{"GENESIS"} value as \texttt{prev\_hash}. This provides tamper-evidence for modification or deletion. As in the reference architecture, confidentiality is out of scope for the PoC store; deployments that log sensitive metadata would additionally require access controls and encryption at rest.

\subsection{\use in the PoC: verification and basic auditor workflow}

\begin{figure}[t]
  \centering
  \includegraphics[width=0.3\linewidth]{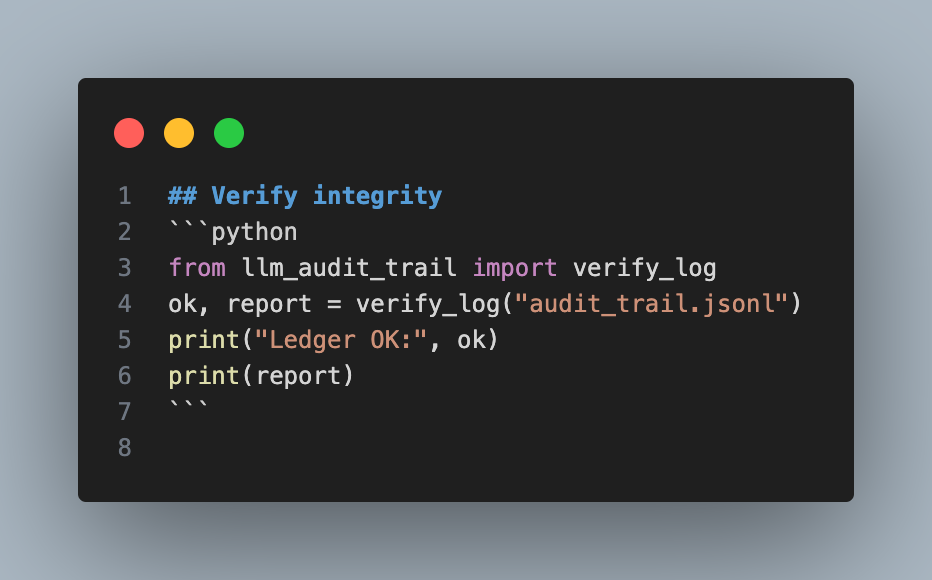}
    \Description{Screenshot Python code for verifying the integrity of an audit log by replaying the hash chain.}
  \caption{Example Python code for verifying the integrity of an audit log by replaying the hash chain.}
  \label{fig:verify-code}
\end{figure}

The PoC \use centers on a verifier function(see Fig. \ref{fig:verify-code}) \texttt{verify\_log(path)} that replays the hash chain and returns a boolean plus a report summarizing the number of events checked and the index of any first mismatch. Combined with simple filters over fields like \texttt{model\_id}, \texttt{dataset\_id}, \texttt{deployment\_id}, and \texttt{event\_type}, an auditor can reconstruct timelines such as all events involving a model between two dates, or all deployment approvals for a given production environment.

\subsection{Integration patterns and walkthrough documentation}
\label{sec:integration}

The PoC repository includes a README that serves as an end-to-end guide to using the audit trail across the main lifecycle touchpoints, including dataset registration, training and evaluation instrumentation, deployment-time logging, governance decision capture, and integrity verification. The README is organized around these integration patterns and shows how each emitter and interface can be adopted independently and then composed into a shared, chronological trail. For concreteness, the repository also includes a small illustrative \texttt{audit\_trail.jsonl} produced by the example components; we include an excerpt in Appendix~\ref{app:audit-excerpt}.

\subsection{Practicality and integration takeaways}

The implementation of \texttt{llm-audit-trail} suggests that an LLM audit trail can be practical even when implemented as a small, reusable layer within existing development and deployment workflows. For capture, training instrumentation can be added as a localized callback and serving instrumentation as middleware, which leaves existing workflows largely unchanged and keeps integration overhead low. For governance, the CLI allows risk and compliance actors to record structured decision events without touching application code, while still writing into the same chronological trail as technical telemetry. On the use side, integrity verification reduces to replaying the hash chain, and simple scoped filters are sufficient to reconstruct basic timelines for review. The implementation is also open source and intentionally modular, so teams can adapt the emitters and governance interfaces to their local LLM stacks and deployment constraints while preserving a shared, verifiable core trail.

\section{Revisiting the motivating scenarios}
\label{sec:revisit-scenarios}

We now return to the two motivating scenarios from Section~\ref{sec:motivation} and sketch how an LLM audit trail, instantiated as in Sections~\ref{sec:architecture} and~\ref{sec:implementation}, would support investigation and governance in each case. 

\paragraph{Financial advice chatbot incident.}
Under the proposed framework, the bank’s audit trail would contain a sequence of events from base model selection through fine tuning, evaluation, deployment, and subsequent configuration changes, all keyed by stable identifiers such as \texttt{model\_id} and \texttt{deployment\_id}. In practice, investigators could query the ledger for all events involving the deployed mortgage assistant around the date of the complaint: \texttt{FineTuneStart/End} and \texttt{Evaluation} events that fix the training data, code version, and performance at sign off, \texttt{DeploymentCompleted} and \texttt{RolloutChanged} events that show when the assistant was exposed to which user segments, and \texttt{ServingConfigChanged} events that record prompt templates, decoding parameters, and retrieval hooks in effect at the time. Approval and risk waiver events created through the governance CLI would show who authorized the initial deployment and any subsequent expansions or exceptions, along with stated rationales or constraints. The bank still needs to analyze logs and interview staff to understand why the specific customer received unsuitable advice, but it no longer needs to reconstruct basic facts about versions, configurations, and approvals from memory, since these elements appear as a verifiable chain of events in the audit trail.

\paragraph{Clinical documentation assistant.}
In the clinical documentation scenario, an audit trail would link foundation or base model choices, fine tuning on local discharge summaries and templates, and EHR specific deployment details into a single chain keyed by \texttt{model\_id}, \texttt{dataset\_id}, and \texttt{deployment\_id}. Vendor side events would record which corpora of notes and structured fields were used for training, how follow up instructions, medication lists, and billing codes were represented in the fine tuning data, and what evaluation criteria were used at sign off, for example rates of omitted follow up actions or coding errors on held out cases. Hospital side events would show when a particular model version was integrated into the discharge workflow at each unit, whether clinicians were required to review and edit drafts, which prompt templates or EHR plug in configurations were active, and what approvals and usage policies governed its use. As concerns emerge about missing follow up instructions or shifts in coding patterns, compliance and quality teams could query the trail to see when prompts or templates were changed, whether guardrails for mandatory sections were enabled, when monitoring jobs last evaluated note completeness against checklists, and how any corrective actions or rollbacks were recorded. The trail would not substitute for clinical chart review or external audit of billing practices, but it would provide a structured, time stamped account of how the documentation assistant was trained, configured, and embedded in workflow at the points when harms are alleged.

\section{Discussion}

Audit trails transform accountability from an aspirational goal into a continuous, operational capability. They do not merely document a model’s existence but preserve its evolution, decisions, and context in a verifiable form. In this section, we discuss the sociotechnical conditions that determine their effectiveness, the challenges of and scalability, and the broader ethical, legal, and institutional questions they raise.

\paragraph{Epistemic limits.}
Audit trails provide chronological and structural traceability but not causal explanation. They verify that events occurred in a verifiable sequence without showing that one directly caused another. Causation and responsibility still depend on contextual interpretation and organizational judgment. 

\paragraph{Scalability and complex system composition.}
Operationalizing audit trails over long-lived, high-throughput systems introduces challenges of scale. Continuous inference, fine-tuning, and monitoring can generate millions of records to process. Accountability requires tooling that can summarize the records and query, filter, and surface anomalies or policy violations automatically. 
Real-world deployments rarely involve a single model: LLMs interact with retrieval modules, ranking systems, and decision layers. Extending audit trails to these multi-component architectures calls for synchronized schemas and event correlation across heterogeneous systems. 

\paragraph{Ethical and institutional tensions.}
Audit trails strengthen transparency but also shift the burden of accountability inward from external watchdogs to developers and deployers. This redistribution raises questions about who controls access to logs, how long data should be retained, and how audit data might itself be misused. Privacy and proportionality are therefore guiding principles. Logs should minimize personal or sensitive content, favor metadata over raw text, and apply structured redaction or aggregation. Equally, the existence of logs should not legitimize surveillance of workers or users. Governance must ensure that access to logs is legitimate, proportionate, and itself auditable.

\section{Conclusion}
LLM audit trails are a reusable layer that turns heterogeneous events from fine tuning, deployment, and monitoring into a common, tamper evident ledger. Our framework specifies what to record and why, our reference architecture shows how to connect emitters, repositories, and auditor interfaces, and our Python proof of concept demonstrates that such a layer can be integrated into existing workflows with modest effort. For developers, the audit layer turns existing MLOps events into a structured, tamper evident ledger with minimal code changes. For auditors and regulators, it offers a starting point for specifying concrete evidence requirements in LLM deployments, rather than relying solely on static documentation. The resulting logs provide shared evidence about how powerful models are constructed and used.

Future research should evaluate audit trails not only as technical artifacts but as sociotechnical infrastructures. Empirical studies could examine how developers, auditors, and regulators actually engage with audit trail data; field experiments could test whether audit availability changes behavior. Technical work could advance cryptographically verifiable, multi-writer coordination at scale, and differential privacy for sensitive logs. Policy and standards efforts could define sectoral “audit profiles” specifying minimal evidence requirements. Over time, such efforts could make auditability a measurable property of trustworthy AI systems on par with accuracy, robustness, and fairness.

\newpage
\bibliographystyle{ACM-Reference-Format}
\bibliography{references}

@misc{hsbc_mistral,
    author = {{HSBC}},
	title = {We’re partnering with {AI} powerhouse {Mistral} {\textbar} {HSBC} news {\textbar} {HSBC} {Holdings} plc},
	howpublished  = {\url{https://www.hsbc.com/news-and-views/news/hsbc-news-archive/we-re-partnering-with-ai-powerhouse-mistral}},
	language = {en},
    year = {2025},
	urldate = {2026-01-02},
	journal = {HSBC},
}

@article{crisanto_regulating_2024,
    journal ={FSI Insights},
	title = {Regulating {AI} in the financial sector: recent developments and main challenges},
	shorttitle = {Regulating {AI} in the financial sector},
	url = {https://www.bis.org/fsi/publ/insights63.htm},
	language = {en},
	urldate = {2026-01-02},
	author = {Crisanto, Juan Carlos and Leuterio, Cris Benson and Prenio, Jermy and Yong, Jeffery},
	month = dec,
	year = {2024},
}

@article{arndt_tethered_2017,
	title = {Tethered to the {EHR}: {Primary} {Care} {Physician} {Workload} {Assessment} {Using} {EHR} {Event} {Log} {Data} and {Time}-{Motion} {Observations}},
	volume = {15},
	issn = {1544-1717},
	shorttitle = {Tethered to the {EHR}},
	doi = {10.1370/afm.2121},
	language = {eng},
	number = {5},
	journal = {Annals of Family Medicine},
	author = {Arndt, Brian G. and Beasley, John W. and Watkinson, Michelle D. and Temte, Jonathan L. and Tuan, Wen-Jan and Sinsky, Christine A. and Gilchrist, Valerie J.},
	month = sep,
	year = {2017},
	pmid = {28893811},
	pmcid = {PMC5593724},
	pages = {419--426},
}

@article{sinsky_allocation_2016,
	title = {Allocation of {Physician} {Time} in {Ambulatory} {Practice}: {A} {Time} and {Motion} {Study} in 4 {Specialties}},
	volume = {165},
	issn = {1539-3704},
	shorttitle = {Allocation of {Physician} {Time} in {Ambulatory} {Practice}},
	doi = {10.7326/M16-0961},
	language = {eng},
	number = {11},
	journal = {Annals of Internal Medicine},
	author = {Sinsky, Christine and Colligan, Lacey and Li, Ling and Prgomet, Mirela and Reynolds, Sam and Goeders, Lindsey and Westbrook, Johanna and Tutty, Michael and Blike, George},
	month = dec,
	year = {2016},
	pmid = {27595430},
	pages = {753--760},
}

@article{10.1001/jamanetworkopen.2025.34982,
    author = {Shah, Kaustav P. and Johnson, Kevin B.},
    title = {The Ambient AI Scribe Revolution—Early Gains and Open Questions},
    journal = {JAMA Network Open},
    volume = {8},
    number = {10},
    pages = {e2534982-e2534982},
    year = {2025},
    month = {10},
    issn = {2574-3805},
    doi = {10.1001/jamanetworkopen.2025.34982},
    url = {https://doi.org/10.1001/jamanetworkopen.2025.34982},
}

@inproceedings{Cobbe2023SupplyChain,
author = {Cobbe, Jennifer and Veale, Michael and Singh, Jatinder},
title = {Understanding accountability in algorithmic supply chains},
year = {2023},
isbn = {9798400701924},
publisher = {Association for Computing Machinery},
address = {New York, NY, USA},
url = {https://doi.org/10.1145/3593013.3594073},
doi = {10.1145/3593013.3594073},
booktitle = {Proceedings of the 2023 ACM Conference on Fairness, Accountability, and Transparency},
pages = {1186–1197},
numpages = {12},
location = {Chicago, IL, USA},
series = {FAccT '23}
}

@inproceedings{10.5555/3666122.3668547,
author = {Schaeffer, Rylan and Miranda, Brando and Koyejo, Sanmi},
title = {Are emergent abilities of large language models a mirage?},
year = {2023},
publisher = {Curran Associates Inc.},
address = {Red Hook, NY, USA},
booktitle = {Proceedings of the 37th International Conference on Neural Information Processing Systems},
articleno = {2425},
numpages = {17},
location = {New Orleans, LA, USA},
series = {NIPS '23}
}

@misc{fiddler_observability_2025,
  title   = {{AI} Observability},
  author  = {{Fiddler AI}},
  year    = {2025},
  url     = {https://www.fiddler.ai/ai-observability}
}

@inproceedings{10.5555/3666122.3669397,
author = {Turpin, Miles and Michael, Julian and Perez, Ethan and Bowman, Samuel R.},
title = {Language models don't always say what they think: unfaithful explanations in chain-of-thought prompting},
year = {2023},
publisher = {Curran Associates Inc.},
address = {Red Hook, NY, USA},
booktitle = {Proceedings of the 37th International Conference on Neural Information Processing Systems},
articleno = {3275},
numpages = {14},
location = {New Orleans, LA, USA},
series = {NIPS '23}
}

@misc{weights_bias,
	title = {Weights \& {Biases}: {The} {AI} {Developer} {Platform}},
	shorttitle = {Weights \& {Biases}},
    author = {Weights and Biases},
	url = {https://wandb.ai/site/},
	language = {en-US},
	date = {2025-11-19},
	journal = {Weights \& Biases},
}

@inproceedings{10.1145/3399579.3399867,
author = {Chen, Andrew and Chow, Andy and Davidson, Aaron and DCunha, Arjun and Ghodsi, Ali and Hong, Sue Ann and Konwinski, Andy and Mewald, Clemens and Murching, Siddharth and Nykodym, Tomas and Ogilvie, Paul and Parkhe, Mani and Singh, Avesh and Xie, Fen and Zaharia, Matei and Zang, Richard and Zheng, Juntai and Zumar, Corey},
title = {Developments in MLflow: A System to Accelerate the Machine Learning Lifecycle},
year = {2020},
isbn = {9781450380232},
publisher = {Association for Computing Machinery},
address = {New York, NY, USA},
url = {https://doi.org/10.1145/3399579.3399867},
doi = {10.1145/3399579.3399867},
booktitle = {Proceedings of the Fourth International Workshop on Data Management for End-to-End Machine Learning},
articleno = {5},
numpages = {4},
location = {Portland, OR, USA},
series = {DEEM '20}
}

@INPROCEEDINGS{10516659,
  author={Birhane, Abeba and Steed, Ryan and Ojewale, Victor and Vecchione, Briana and Raji, Inioluwa Deborah},
  booktitle={2024 IEEE Conference on Secure and Trustworthy Machine Learning (SaTML)}, 
  title={AI auditing: The Broken Bus on the Road to AI Accountability}, 
  year={2024},
  volume={},
  number={},
  pages={612-643},
  doi={10.1109/SaTML59370.2024.00037}}

@article{lai_large_2024,
	title = {Large language models in law: {A} survey},
	volume = {5},
	issn = {2666-6510},
	shorttitle = {Large language models in law},
	url = {https://www.sciencedirect.com/science/article/pii/S2666651024000172},
	doi = {10.1016/j.aiopen.2024.09.002},
	urldate = {2025-01-21},
	journal = {AI Open},
	author = {Lai, Jinqi and Gan, Wensheng and Wu, Jiayang and Qi, Zhenlian and Yu, Philip S.},
	month = jan,
	year = {2024},
	pages = {181--196},
}

@misc{nazi_large_2024,
	title = {Large language models in healthcare and medical domain: {A} review},
	shorttitle = {Large language models in healthcare and medical domain},
	url = {http://arxiv.org/abs/2401.06775},
	doi = {10.48550/arXiv.2401.06775},
	urldate = {2025-01-21},
	publisher = {arXiv},
	author = {Nazi, Zabir Al and Peng, Wei},
	month = jul,
	year = {2024},
	note = {arXiv:2401.06775 [cs]},
}

@inproceedings{li_large_2023,
	address = {New York, NY, USA},
	series = {{ICAIF} '23},
	title = {Large {Language} {Models} in {Finance}: {A} {Survey}},
	isbn = {9798400702402},
	shorttitle = {Large {Language} {Models} in {Finance}},
	url = {https://dl.acm.org/doi/10.1145/3604237.3626869},
	doi = {10.1145/3604237.3626869},
	urldate = {2025-01-21},
	booktitle = {Proceedings of the {Fourth} {ACM} {International} {Conference} on {AI} in {Finance}},
	publisher = {Association for Computing Machinery},
	author = {Li, Yinheng and Wang, Shaofei and Ding, Han and Chen, Hang},
	month = nov,
	year = {2023},
	pages = {374--382},
}

@article{Arnold2019,
  author       = {Matthew Arnold and Rachel K. E. Bellamy and Michael Hind and Stephanie Houde and Sameep Mehta and Aleksandra Mojsilovi{\'c} and Ravi Nair and Karthikeyan Natesan~Ramamurthy and Alexandra Olteanu and David Piorkowski and Darrell Reimer and John T. Richards and Jason Tsay and Kush R. Varshney},
  title        = {{FactSheets}: Increasing Trust in {AI} Services through Supplier’s Declarations of Conformity},
  journal      = {{IBM} Journal of Research and Development},
  volume       = {63},
  number       = {4/5},
  pages        = {6:1--6:13},
  year         = {2019},
  doi          = {10.1147/JRD.2019.2942288},
  url          = {https://doi.org/10.1147/JRD.2019.2942288}
}

@article{Mokander2023,
	title = {Auditing large language models: a three-layered approach},
	volume = {4},
	issn = {2730-5961},
	shorttitle = {Auditing large language models},
	url = {https://doi.org/10.1007/s43681-023-00289-2},
	doi = {10.1007/s43681-023-00289-2},
	language = {en},
	number = {4},
	urldate = {2025-05-14},
	journal = {AI and Ethics},
	author = {Mökander, Jakob and Schuett, Jonas and Kirk, Hannah Rose and Floridi, Luciano},
	month = nov,
	year = {2024},
	pages = {1085--1115},
}

@inproceedings{Mitchell2019,
  author       = {Margaret Mitchell and Simone Wu and Andrew Zaldivar and Parker Barnes and Lucy Vasserman and Ben Hutchinson and Elena Spitzer and Inioluwa~Deborah Raji and Timnit Gebru},
  title        = {{Model Cards for Model Reporting}},
  booktitle    = {Proceedings of the Conference on Fairness, Accountability, and Transparency (FAT*~'19)},
  pages        = {220--229},
  year         = {2019},
  publisher    = {ACM},
  address      = {New York, NY, USA},
  doi          = {10.1145/3287560.3287596},
  url          = {https://doi.org/10.1145/3287560.3287596}
}

@article{Gebru2018,
  author       = {Timnit Gebru and Jamie Morgenstern and Briana Vecchione and Jennifer~Wortman Vaughan and Hanna~M. Wallach and Hal Daum{\'e}~III and Kate Crawford},
  title        = {{Datasheets for Datasets}},
  journal      = {Communications of the ACM},
  volume       = {64},
  number       = {12},
  pages        = {86--92},
  year         = {2021},
  doi          = {10.1145/3458723},
  url          = {https://doi.org/10.1145/3458723}
}

@misc{mattu_how_nodate,
	title = {How {We} {Analyzed} the {COMPAS} {Recidivism} {Algorithm}},
	url = {https://www.propublica.org/article/how-we-analyzed-the-compas-recidivism-algorithm},
	language = {en},
	   year  = 2016,
	journal = {ProPublica},
	author = {Julia Angwin and Lauren Kirchner and Surya Mattu and Jeff Larson},
}

@article{Widder2023Modularity,
	title = {Dislocated accountabilities in the  “{AI} supply chain”: {Modularity} and developers’ notions of responsibility},
	volume = {10},
	issn = {2053-9517},
	shorttitle = {Dislocated accountabilities in the  “{AI} supply chain”},
	url = {https://doi.org/10.1177/20539517231177620},
	doi = {10.1177/20539517231177620},
	language = {en},
	number = {1},
	urldate = {2024-12-07},
	journal = {Big Data \& Society},
	author = {Widder, David Gray and Nafus, Dawn},
	month = jan,
	year = {2023},
	note = {Publisher: SAGE Publications Ltd},
	pages = {20539517231177620},
}

@article{Thompson1980Responsibility,
	title = {Moral {Responsibility} of {Public} {Officials}: {The} {Problem} of {Many} {Hands}},
	volume = {74},
	issn = {0003-0554, 1537-5943},
	shorttitle = {Moral {Responsibility} of {Public} {Officials}},
	url = {https://www.cambridge.org/core/journals/american-political-science-review/article/abs/moral-responsibility-of-public-officials-the-problem-of-many-hands/39DD3FAB7BF7DC7A242407143674F22B},
	doi = {10.2307/1954312},
	language = {en},
	number = {4},
	urldate = {2024-12-07},
	journal = {American Political Science Review},
	author = {Thompson, Dennis F.},
	month = dec,
	year = {1980},
	pages = {905--916},
}

@article{Bommasani2024Ecosystem,
	title = {Ecosystem {Graphs}: {Documenting} the {Foundation} {Model} {Supply} {Chain}},
	volume = {7},
	copyright = {Copyright (c) 2024 Association for the Advancement of Artificial Intelligence},
	issn = {3065-8365},
	shorttitle = {Ecosystem {Graphs}},
	url = {https://ojs.aaai.org/index.php/AIES/article/view/31629},
	doi = {10.1609/aies.v7i1.31629},
	language = {en},
	urldate = {2024-12-07},
	journal = {Proceedings of the AAAI/ACM Conference on AI, Ethics, and Society},
	author = {Bommasani, Rishi and Soylu, Dilara and Liao, Thomas I. and Creel, Kathleen A. and Liang, Percy},
	month = oct,
	year = {2024},
	pages = {196--209},
}

@misc{hopkins2025aisupplychainsemerging,
      title={AI Supply Chains: An Emerging Ecosystem of AI Actors, Products, and Services}, 
      author={Aspen Hopkins and Sarah H. Cen and Andrew Ilyas and Isabella Struckman and Luis Videgaray and Aleksander Mądry},
      year={2025},
      eprint={2504.20185},
      archivePrefix={arXiv},
      primaryClass={cs.CY},
      url={https://arxiv.org/abs/2504.20185}, 
}

@techreport{alsallakh2022system,
  title={System-level transparency of machine learning},
  author={Alsallakh, Bilal and Cheema, Adeel and Procope, Chavez and Adkins, David and McReynolds, Emily and Wang, Erin and Pehl, Grace and Green, Nekesha and Zvyagina, Polina},
  year={2022},
  institution={Technical Report}
}

@misc{euaiact,
    url = "http://data.europa.eu/eli/reg/2024/1689/oj",
    title = {Regulation (EU) 2024/1689 of the European Parliament and of the Council of 13 June 2024 laying down harmonised rules on artificial intelligence and amending Regulations (EC) No 300/2008, (EU) No 167/2013, (EU) No 168/2013, (EU) 2018/858, (EU) 2018/1139 and (EU) 2019/2144 and Directives 2014/90/EU, (EU) 2016/797 and (EU) 2020/1828 (Artificial Intelligence Act) (Text with EEA relevance)},
    number = {1689},
    author = {Council of the European Union and European Parliament},
    year = {2024},
    month = {Jul}
}

@inproceedings{buolamwini_gender_2018,
	title = {Gender {Shades}: {Intersectional} {Accuracy} {Disparities} in {Commercial} {Gender} {Classification}},
	shorttitle = {Gender {Shades}},
	url = {https://proceedings.mlr.press/v81/buolamwini18a.html},
	language = {en},
	urldate = {2025-05-14},
	booktitle = {Proceedings of the 1st {Conference} on {Fairness}, {Accountability} and {Transparency}},
	publisher = {PMLR},
	author = {Buolamwini, Joy and Gebru, Timnit},
	month = jan,
	year = {2018},
	note = {ISSN: 2640-3498},
	pages = {77--91},
}

@misc{ouyang2022,
      title={Training language models to follow instructions with human feedback}, 
      author={Long Ouyang and Jeff Wu and Xu Jiang and Diogo Almeida and Carroll L. Wainwright and Pamela Mishkin and Chong Zhang and Sandhini Agarwal and Katarina Slama and Alex Ray and John Schulman and Jacob Hilton and Fraser Kelton and Luke Miller and Maddie Simens and Amanda Askell and Peter Welinder and Paul Christiano and Jan Leike and Ryan Lowe},
      year={2022},
      eprint={2203.02155},
      archivePrefix={arXiv},
      primaryClass={cs.CL},
      url={https://arxiv.org/abs/2203.02155}, 
}

@ARTICLE{mlops2023,
  author={Kreuzberger, Dominik and Kühl, Niklas and Hirschl, Sebastian},
  journal={IEEE Access}, 
  title={Machine Learning Operations (MLOps): Overview, Definition, and Architecture}, 
  year={2023},
  volume={11},
  number={},
  pages={31866-31879},
  doi={10.1109/ACCESS.2023.3262138}}

@inproceedings{Vartak2016,
	address = {San Francisco California},
	title = {ModelDB: a system for machine learning model management},
	isbn = {978-1-4503-4207-0},
	shorttitle = {M {\textless}span style="font-variant},
	url = {https://dl.acm.org/doi/10.1145/2939502.2939516},
	doi = {10.1145/2939502.2939516},
	language = {en},
	urldate = {2025-05-14},
	booktitle = {Proceedings of the {Workshop} on {Human}-{In}-the-{Loop} {Data} {Analytics}},
	publisher = {ACM},
	author = {Vartak, Manasi and Subramanyam, Harihar and Lee, Wei-En and Viswanathan, Srinidhi and Husnoo, Saadiyah and Madden, Samuel and Zaharia, Matei},
	month = jun,
	year = {2016},
	pages = {1--3},
}

@article{Diakopoulos2016,
author = {Diakopoulos, Nicholas},
title = {Accountability in algorithmic decision making},
year = {2016},
issue_date = {February 2016},
publisher = {Association for Computing Machinery},
address = {New York, NY, USA},
volume = {59},
number = {2},
issn = {0001-0782},
url = {https://doi.org/10.1145/2844110},
doi = {10.1145/2844110},
journal = {Commun. ACM},
month = jan,
pages = {56–62},
numpages = {7}
}

@inproceedings{10.1145/3442188.3445921,
author = {Cobbe, Jennifer and Lee, Michelle Seng Ah and Singh, Jatinder},
title = {Reviewable Automated Decision-Making: A Framework for Accountable Algorithmic Systems},
year = {2021},
isbn = {9781450383097},
publisher = {Association for Computing Machinery},
address = {New York, NY, USA},
url = {https://doi.org/10.1145/3442188.3445921},
doi = {10.1145/3442188.3445921},
booktitle = {Proceedings of the 2021 ACM Conference on Fairness, Accountability, and Transparency},
pages = {598–609},
numpages = {12},
location = {Virtual Event, Canada},
series = {FAccT '21}
}

@inproceedings{10.1145/3531146.3534628,
author = {Lima, Gabriel and Grgi\'{c}-Hla\v{c}a, Nina and Jeong, Jin Keun and Cha, Meeyoung},
title = {The Conflict Between Explainable and Accountable Decision-Making Algorithms},
year = {2022},
isbn = {9781450393522},
publisher = {Association for Computing Machinery},
address = {New York, NY, USA},
url = {https://doi.org/10.1145/3531146.3534628},
doi = {10.1145/3531146.3534628},
booktitle = {Proceedings of the 2022 ACM Conference on Fairness, Accountability, and Transparency},
pages = {2103–2113},
numpages = {11},
location = {Seoul, Republic of Korea},
series = {FAccT '22}
}

@inproceedings{10.1145/3706598.3713301,
author = {Ojewale, Victor and Steed, Ryan and Vecchione, Briana and Birhane, Abeba and Raji, Inioluwa Deborah},
title = {Towards AI Accountability Infrastructure: Gaps and Opportunities in AI Audit Tooling},
year = {2025},
isbn = {9798400713941},
publisher = {Association for Computing Machinery},
address = {New York, NY, USA},
url = {https://doi.org/10.1145/3706598.3713301},
doi = {10.1145/3706598.3713301},
booktitle = {Proceedings of the 2025 CHI Conference on Human Factors in Computing Systems},
articleno = {815},
numpages = {29},
location = {
},
series = {CHI '25}
}

@misc{AWSGovernance,
    author = {{AWS}},
	title = {{ML} {Governance} with {Amazon} {SageMaker}},
	url = {https://aws.amazon.com/sagemaker/ai/ml-governance/},
	language = {en-US},
	urldate = {2025-09-30},
	journal = {Amazon Web Services, Inc.},
}

@misc{DataRobot2022,
    author = {DataRobot},
	title = {{AI} {Governance}},
	url = {https://www.datarobot.com/product/ai-governance/},
	language = {en-US},
    year = 2025,
	urldate = {2025-09-30},
	journal = {DataRobot},
}

@inproceedings{bender2021,
author = {Bender, Emily M. and Gebru, Timnit and McMillan-Major, Angelina and Shmitchell, Shmargaret},
title = {On the Dangers of Stochastic Parrots: Can Language Models Be Too Big?},
year = {2021},
isbn = {9781450383097},
publisher = {Association for Computing Machinery},
address = {New York, NY, USA},
url = {https://doi.org/10.1145/3442188.3445922},
doi = {10.1145/3442188.3445922},
booktitle = {Proceedings of the 2021 ACM Conference on Fairness, Accountability, and Transparency},
pages = {610–623},
numpages = {14},
location = {Virtual Event, Canada},
series = {FAccT '21}
}

@inproceedings{kroll2021,
author = {Kroll, Joshua A.},
title = {Outlining Traceability: A Principle for Operationalizing Accountability in Computing Systems},
year = {2021},
isbn = {9781450383097},
publisher = {Association for Computing Machinery},
address = {New York, NY, USA},
url = {https://doi.org/10.1145/3442188.3445937},
doi = {10.1145/3442188.3445937},
booktitle = {Proceedings of the 2021 ACM Conference on Fairness, Accountability, and Transparency},
pages = {758–771},
numpages = {14},
location = {Virtual Event, Canada},
series = {FAccT '21}
}

@article{Felzmann2020,
	title = {Towards {Transparency} by {Design} for {Artificial} {Intelligence}},
	volume = {26},
	issn = {1471-5546},
	doi = {10.1007/s11948-020-00276-4},
	language = {eng},
	number = {6},
	journal = {Science and Engineering Ethics},
	author = {Felzmann, Heike and Fosch-Villaronga, Eduard and Lutz, Christoph and Tamò-Larrieux, Aurelia},
	month = dec,
	year = {2020},
	pmid = {33196975},
	pmcid = {PMC7755865},
	pages = {3333--3361},
}

@article{
Obermeyer2019,
author = {Ziad Obermeyer  and Brian Powers  and Christine Vogeli  and Sendhil Mullainathan },
title = {Dissecting racial bias in an algorithm used to manage the health of populations},
journal = {Science},
volume = {366},
number = {6464},
pages = {447-453},
year = {2019},
doi = {10.1126/science.aax2342},
URL = {https://www.science.org/doi/abs/10.1126/science.aax2342},
eprint = {https://www.science.org/doi/pdf/10.1126/science.aax2342}}

@inproceedings{10.1145/3351095.3372833,
author = {Wieringa, Maranke},
title = {What to account for when accounting for algorithms: a systematic literature review on algorithmic accountability},
year = {2020},
isbn = {9781450369367},
publisher = {Association for Computing Machinery},
address = {New York, NY, USA},
url = {https://doi.org/10.1145/3351095.3372833},
doi = {10.1145/3351095.3372833},
booktitle = {Proceedings of the 2020 Conference on Fairness, Accountability, and Transparency},
pages = {1–18},
numpages = {18},
location = {Barcelona, Spain},
series = {FAT* '20}
}

@techreport{NIST2023,
  author       = {Elham Tabassi},
  title        = {{Artificial Intelligence Risk Management Framework (AI RMF 1.0)}},
  institution  = {National Institute of Standards and Technology},
  address      = {Gaithersburg, MD},
  number       = {NIST AI 100-1},
  year         = {2023},
  note         = {NIST Trustworthy and Responsible AI},
  doi          = {10.6028/NIST.AI.100-1},
  url          = {https://doi.org/10.6028/NIST.AI.100-1}
}

@article{Liang2023HELM,
  author       = {Percy Liang and Rishi Bommasani and Tony Lee and Dimitris Tsipras and Dilara Soylu and Michihiro Yasunaga and Yian Zhang and Deepak Narayanan and Yuhuai Wu and Ananya Kumar and Benjamin Newman and Binhang Yuan and Bobby Yan and Ce Zhang and Christian Cosgrove and Christopher~D. Manning and Christopher R{\'e} and Diana Acosta-Navas and Drew~A. Hudson and Eric Zelikman and Esin Durmu{\c{s}} and Faisal Ladhak and Frieda Rong and Hongyu Ren and Huaxiu Yao and Jue Wang and Keshav Santhanam and Laurel Orr and Lucia Zheng and Mert Yuksekgonul and Mirac Suzgun and Nathan Kim and Neel Guha and Niladri~S. Chatterji and Omar Khattab and Peter Henderson and Qian Huang and Ryan~A. Chi and Sang~Michael Xie and Shibani Santurkar and Surya Ganguli and Tatsunori Hashimoto and Thomas Icard and Tianyi Zhang and Vishrav Chaudhary and William~Yang Wang and Xuechen Li and Yifan Mai and Yuhui Zhang and Yuta Koreeda},
  title        = {{Holistic Evaluation of Language Models}},
  journal      = {Transactions on Machine Learning Research},
  year         = {2023},
  note         = {OpenReview TMLR (published Jan 2023)},
  url          = {https://openreview.net/forum?id=iO4LZibEqW}
}

@inproceedings{Raji2020,
  author       = {Inioluwa~Deborah Raji and Andrew Smart and Rebecca~N. White and Margaret Mitchell and Timnit Gebru and Ben Hutchinson and Jamila Smith-Loud and Daniel Theron and Parker Barnes},
  title        = {{Closing the AI Accountability Gap: Defining an End-to-End Framework for Internal Algorithmic Auditing}},
  booktitle    = {Proceedings of the 2020 Conference on Fairness, Accountability, and Transparency (FAT*~'20)},
  pages        = {33--44},
  year         = {2020},
  publisher    = {ACM},
  address      = {New York, NY, USA},
  doi          = {10.1145/3351095.3372873},
  url          = {https://doi.org/10.1145/3351095.3372873}
}

@article{Bommasani2021FoundationModels,
  author       = {Rishi Bommasani and Drew~A. Hudson and Ehsan Adeli and Russ Altman and Simran Arora and Sydney von Arx and Michael~S. Bernstein and Jeannette Bohg and Antoine Bosselut and Emma Brunskill and Erik Brynjolfsson and Shyamal Buch and Dallas Card and Rodrigo Castellon and Niladri~S. Chatterji and Annie~S. Chen and Kathleen Creel and Jared~Q. Davis and Dorottya Demszky and Chris Donahue and Moussa Doumbouya and Esin Durmu{\c{s}} and Stefano Ermon and John Etchemendy and Kawin Ethayarajh and Li Fei-Fei and Chelsea Finn and Trevor Gale and Lauren~E. Gillespie and Karan Goel and Noah~D. Goodman and Shelby Grossman and Neel Guha and Tatsunori Hashimoto and Peter Henderson and John Hewitt and Daniel~E. Ho and Jenny Hong and Kyle Hsu and Jing Huang and Thomas Icard and Saahil Jain and Dan Jurafsky and Pratyusha Kalluri and Siddharth Karamcheti and Geoff Keeling and Fereshte Khani and Omar Khattab and Pang~Wei Koh and Mark~S. Krass and Ranjay Krishna and Rohith Kuditipudi and et al.},
  title        = {{On the Opportunities and Risks of Foundation Models}},
  journal      = {arXiv e-prints},
  volume       = {arXiv:2108.07258},
  year         = {2021},
  url          = {https://arxiv.org/abs/2108.07258}
}

@misc{groeneveld2024olmoacceleratingsciencelanguage,
      title={OLMo: Accelerating the Science of Language Models}, 
      author={Dirk Groeneveld and Iz Beltagy and Pete Walsh and Akshita Bhagia and Rodney Kinney and Oyvind Tafjord and Ananya Harsh Jha and Hamish Ivison and Ian Magnusson and Yizhong Wang and Shane Arora and David Atkinson and Russell Authur and Khyathi Raghavi Chandu and Arman Cohan and Jennifer Dumas and Yanai Elazar and Yuling Gu and Jack Hessel and Tushar Khot and William Merrill and Jacob Morrison and Niklas Muennighoff and Aakanksha Naik and Crystal Nam and Matthew E. Peters and Valentina Pyatkin and Abhilasha Ravichander and Dustin Schwenk and Saurabh Shah and Will Smith and Emma Strubell and Nishant Subramani and Mitchell Wortsman and Pradeep Dasigi and Nathan Lambert and Kyle Richardson and Luke Zettlemoyer and Jesse Dodge and Kyle Lo and Luca Soldaini and Noah A. Smith and Hannaneh Hajishirzi},
      year={2024},
      eprint={2402.00838},
      archivePrefix={arXiv},
      primaryClass={cs.CL},
      url={https://arxiv.org/abs/2402.00838}, 
}

\appendix

\newpage
\section{Illustrative audit trail excerpt}
\label{app:audit-excerpt}

To make the PoC output concrete, we show a short excerpt from the newline-delimited JSON log produced by the example components in the repository. The ledger interleaves dataset, training, evaluation, governance, and serving metadata as a single chronological stream, with each record carrying scoped identifiers and integrity fields that support later verification.

\begin{lstlisting}[basicstyle=\ttfamily\small,breaklines=true]
{"event_id":"23ea25...",
 "timestamp":"2025-10-02T18:33:11Z",
 "system":"data_engineering",
 "actor":"Data Eng",
 "event_type":"DatasetRegistered",
 "dataset_id":"hf:stanfordnlp/imdb",
 "model_id":null,
 "deployment_id":null,
 "details":{"source":"huggingface://datasets/stanfordnlp/imdb",
            "version":"latest",
            "rows":100000,
            "license":"unknown"},
 "prev_hash":"7eba94...47f56",
 "curr_hash":"f46658...482a0"}
\end{lstlisting}

The full \texttt{audit\_trail.jsonl} file, along with the README walkthrough describing how to use the library across each integration point, is included in the accompanying codebase.

\end{document}